\documentclass[12pt]{article}

\usepackage{epsfig}
\usepackage{graphicx}
\usepackage{amssymb,amsmath}

\usepackage{epsfig,cite}
\usepackage{color}
\usepackage{graphics}
\usepackage{amsfonts}
\usepackage{epsfig}
\def\one{{\hbox{1\kern-.8mm l}}}

\def\t{{\rm t}}

\def\vp{\varphi }

\def\non{\nonumber}
\def\g{\gamma }

\newcommand{\beq}{\begin{equation}}
\newcommand{\eeq}{\end{equation}}
\newcommand{\be}{\begin{eqnarray}}
 \newcommand{\ee}{\end{eqnarray}}
 \newcommand{\half}{\frac{1}{2}}

\newcommand{\p }{\partial }

\def\a{\alpha }

\def\ep{\epsilon}

\def\g{\gamma }
\def\non{\nonumber}
\def\vp{\varphi}

\def\appendix#1{
  \addtocounter{section}{1}
  \setcounter{equation}{0}
  \renewcommand{\thesection}{\Alph{section}}
  \section*{Appendix \thesection\protect\indent \parbox[t]{11.15cm}
  {#1} }
  \addcontentsline{toc}{section}{Appendix \thesection\ \ \ #1}
  }

\begin{document}
\null\vskip-24pt 
\hfill
UB-ECM-PF-05/10
\vskip-1pt
\hfill
\vskip-1pt
\hfill {\tt hep-th/0504187}
\vskip0.2truecm
\begin{center}
\vskip 0.2truecm {\Large\bf Strong Magnetic Limit of String Theory}

\vskip 2truecm

{\bf  Jorge G. Russo 
}

\bigskip
\medskip

Instituci\' o Catalana de Recerca i Estudis Avan\c{c}ats (ICREA),\\
Departament ECM,
Facultat de F\'\i sica, Universitat de Barcelona,  Spain.

\bigskip

{\tt {jrusso@ecm.ub.es}
}

\end{center}
\begin{abstract}
We show that there exists a certain limit in type I  and type II
superstring theory in the presence of a suitable configuration of
magnetic $U(1)$ fields where all string excitations
get an infinite mass, except for the neutral massless sector and
for the boson and fermion string states lying on the leading Regge
trajectory. For a supersymmetric configuration of magnetic
fields in internal directions,  the resulting theory after the
limit is a 3+1 Lorentz invariant supersymmetric theory. 
Supersymmetry can be broken by introducing
 extra components of the magnetic field or else by finite temperature.
In both cases we compute the one-loop partition function for the type I string model
after taking the limit, which turns out to be different from the
Yang-Mills result that arises by a direct $\a'\to 0$ limit.
 In the case of finite 
temperature, no Hagedorn transition
appears, in consistency with the reduction of the string spectrum.
In type II superstring theory, the analogous limit is constructed by
starting with
a configuration of Melvin twists in two or more complex planes.
The resulting theory contains gravitation plus an infinite number of
 states of the leading Regge trajectory.

\end{abstract}

\date{April 2005}

\vfill\eject




\section{Introduction}
\setcounter{equation}{0}

Taking limits in string theory have led to insightful connections.
One important limit is the low energy limit represented by
$\a'\to 0$, leading to Yang-Mills theory and supergravity \cite{Green}.
In this limit, all superstring excitations become infinitely massive
and decouple, leaving only the massless multiplet.
Another important limit is the one connecting conformal field theories
to string theory in AdS backgrounds
\cite{Maldacena}. The limit is again a low energy $\a'\to 0$ limit,
but this time massive string excitations remain due to the fact that
the limit is taken in a way that one is simultaneously approaching the
horizon region where there is an infinite redshift, leading to  
string excitations of finite mass. 
Furthermore, by taking a suitable limit
of string theory it was shown in \cite{SW} that non-commutative field theories
 exist as quantum field theories.

In the present work we will show that there exists an $\a'\to 0$ limit 
where  string theory gets greatly simplified, due to the fact
that most massive excitations decouple.
The resulting theory includes in the spectrum Yang-Mills vector
fields (in the type I case),
the supergravity multiplet (in the type II case),  
and an infinite tower of states corresponding to
string states lying on the leading Regge trajectory.
The number of states surviving at each mass level $N$ is proportional
to $N$. This is a very small number, compared to the exponential
$\sim \exp[{\rm const.}\sqrt{N}]$ number of states at level $N$ in the
usual string theory spectrum.

Preserving a part of the string spectrum in the limit $\a'\to 0$ 
requires adding
interactions which can lower the energy of a string state
up to zero.
This can be achieved by introducing magnetic fields in some directions.
Due to the gyromagnetic coupling, the magnetic field
lowers the energy of a string state with the spin aligned with the magnetic
field direction (open strings in magnetic fields have been first considered in
 \cite{fradkin,Abou,Nesterenko}). 
This typically leads to  Nielsen-Olesen \cite{nielsen} tachyons for critical values of the magnetic
fields \cite{ferrara,russo}
(similar tachyons in closed string magnetic models are discussed in
\cite{RT1,RT2} and references therein).
Here we shall avoid tachyons by starting with a supersymmetric configuration
of magnetic fields.
In this case, we will see that at a critical magnetic field
(which is infinite in the type I string model and finite in the type II string
model) the states of the leading Regge trajectory with spins aligned
with the magnetic field become massless.
By taking a simultaneous limit with $\a'\to 0$, so that the masses of
these states remain finite and non-zero,
 the masses of all other string states go to infinity, except for the
 neutral massless string states, which remain massless.

This paper is organized as follows.
The models related to type I superstring theory are constructed in section 2.
In section 3 we study the partition function for  non-supersymmetric
models and
for the supersymmetric model at finite temperature.
In the case of finite temperature, 
one finds a finite expression for all $\beta $, which in particular
shows the absence of
a Hagedorn transition.
Finally, in section 4 we construct the type II model.
We show that the resulting theory after the limit contains
the type II massless supergravity multiplet plus and
infinite number of  states with maximum angular momentum.

\section{The string model}
\setcounter{equation}{0}

\subsection{Superstring spectrum and limit}

We first consider type I superstring theory in the presence of two magnetic fields
$F_{45}$ and $F_{67}$.\footnote{In realistic compactifications, the directions 4,5,6,7,8,9 can be compact
so that the string model is Lorentz invariant in 3+1 dimensions.
Models with supersymmetry breaking due to internal magnetic fields 
were constructed in \cite{bachas}.
} The superstring spectrum is given by
\beq
\a' M^ 2= \hat N-\vp_1 \hat J_1 -\vp_2 \hat J_2\ ,
\label{uno}
\eeq
\beq
\hat J_1=  S_{1} -l_1-\half \ ,\ \ \ \ 
\hat J_2=  S_{2} -l_2-\half\ ,
\label{dos}
\eeq
\be
\pi \varphi_1 &=& \arctan 2\pi\a' e F_{45}+\arctan 2\pi\a' e'
F_{45}\ ,\
\non\\
\pi \varphi_2 &=& 
\arctan 2\pi\a'e F_{67}+\arctan 2\pi\a' e' F_{67}\ ,\ 
\label{tres}
\ee
$$
0<\varphi_{1,2}<1\ .
$$
The indices 1 and 2 refer to the directions 45 and 67 respectively, whereas
$e$ and $e'$ are the electric $U(1)$ charges at the two endpoints of
the open string.
Here $\hat N$ includes the normal ordering constant, so that 
$\hat N=0,1,2,...$. The Landau numbers take value $l_{1,2}=0,1,2,...$. The spin quantum numbers of a given state obey the inequality
\beq
| S_1\pm  S_2|\leq \hat N+1\ .
\eeq

{} For generic values of $\vp_1 $ and $\vp_2$, the spectrum exhibits  
\cite{ferrara,russo} magnetic instabilities of Nielsen-Olesen type 
\cite{nielsen}.
States with maximum angular momentum aligned with the magnetic field
become tachyonic above some
critical value of the magnetic field.
As an example, consider the following state (in the light-cone gauge):
\beq
|\Phi_0\rangle =(\alpha_{-1}^{(4)}+i\alpha_{-1}^{(5)})^n 
(b_{-1/2}^{(4)}+ib_{-1/2}^{(5)})\ |0\rangle _{\rm NS}\ ,
\eeq
where $\alpha_n^\mu $ are as usual the mode operators of $X^\mu $, and
$b_r^\mu$ are the mode operators of $\psi^\mu $ in the NS sector.
This state has mass
\beq
\a' M^ 2=n (1-\vp_1)-{(\vp_1-\vp_2)\over 2}\ .
\eeq
It becomes tachyonic for 
all $n$ with  $n<(\vp_1-\vp_2)/(2(1-\vp_1))$,  $\vp_1 >\vp_2$.

If $\vp_1=\vp_2\equiv \vp$, all states have $M^2\geq 0$.
The reason is that this special configuration 
is supersymmetric (preserving 1/2 of the original 16 supersymmetries).
Now consider this configuration and states with $l_1=l_2=0$, 
$ S_1+ S_2=\hat N+1$, so that
$\hat J_1+\hat J_2=\hat N$.
For such states, the mass spectrum (\ref{uno})   becomes
\beq
\a'M^ 2=\hat N (1-\vp )\ .
\label{afa}
\eeq
Next, we write $\a'=\ep \a'_{\rm eff} $ and $\vp=1 -\ep $, and take the
limit $\ep\to 0$. This limit corresponds to sending 
$\a' F_{45}=\a' F_{67}$ to infinity,
$$
2\pi\alpha' F_{45}\to {e+e'\over \pi ee'}\, {1\over \ep }\ .
$$
The limit is similar to the limit of \cite{SW} that leads to
non-commutative super Yang-Mills theory. In the present case,
it is important that there are two components $ F_{45}= F_{67}$.
Also, the open string of \cite{SW} has vanishing total charge, 
$e+e'=0$, whereas
here we consider both sectors, charged and neutral open strings.

The mass spectrum (\ref{afa})
for the states with $ \hat J_1+ \hat J_2=\hat N$ reduces to
\beq
\a'_{\rm eff} M^ 2=\hat N \ ,
\eeq
whereas for any other state in the spectrum one has
$\hat J_1+\hat J_2<\hat N$
so that
\beq
\a'_{\rm eff} M_{\rm other}^ 2={1\over\ep} (\hat N -\hat J_1-\hat
J_2)\ \to\ \infty\ .
\eeq
Thus in the  simultaneous limit $\vp\to 1 $ and $\a'\to 0$,
the states with maximum angular momentum aligned with the magnetic
field
remain with finite mass and
all other string states in the spectrum decouple.

Consider now the neutral sector with $e+e'=0$.
Here the mass spectrum is not affected by the magnetic field,
\beq
\a' M^2=\hat N\ .
\eeq
In the limit $\ep\to 0$, all states are decoupled except for 
the massless states with $\hat N=0$. 
Altogether, the full theory after the limit
contains a massless gauge supermultiplet which is
neutral under the $U(1)$ $F$ field turned on in the 45 and 67
directions,
plus an infinite number of states of the leading Regge trajectory.
This infinite set of states also includes charged vector fields
corresponding
to $\hat N=0$ level having $ S_1=1$, $ S_2=0$ or $ S_1=0$, 
$ S_2=1$.

Let us explicitly construct the surviving charged states.
Define
\be
a_{45}^\dagger &=& \alpha_{-1}^{(4)}+i\alpha_{-1}^{(5)}\ ,\ \ \ \ 
a_{67}^\dagger =\alpha_{-1}^{(6)}+i\alpha_{-1}^{(7)}\ ,
\nonumber\\
b_{-r}^{45} &=& b_{-r}^{(4)}+ib_{-r}^{(5)}\ ,\ \ \ \
b_{-r}^{67} = b_{-r}^{(6)}+ib_{-r}^{(7)}\ ,\ \ \ \ 
b_{-1/2,\pm }^{89}= b_{-1/2}^{(8)}\pm ib_{-1/2}^{(9)}\ ,
\nonumber\\
d_{-n}^{45} &=& d_{-n}^{(4)}+id_{-n}^{(5)}\ ,\ \ \ \
d_{-n}^{67} = d_{-n}^{(6)}+id_{-n}^{(7)}\ ,
\ee
where
$d_n^\mu$ are the mode operators of $\psi^\mu $ in the R sector.
Then the surviving charged states are as follows:

\smallskip

\noindent NS sector:
\be
|\Phi_{\rm boson}^I\rangle  &=& (a_{45}^\dagger)^{n_1} 
(a_{67}^\dagger)^{n_2} b_{-1/2}^{a}\ |0\rangle _{\rm NS}\ ,\qquad
\ \ a=45, \ 67\ ,\ \  n_1+n_2=\hat N\ ,
\nonumber\\
|\Phi_{\rm boson}^{II}\rangle  &=& (a_{45}^\dagger)^{n_1} 
(a_{67}^\dagger)^{n_2} b_{-1/2}^{45} b_{-1/2}^{67}
b_{-1/2,\pm }^{89}
\ |0\rangle _{\rm NS}\ ,\qquad n_1+n_2=\hat N-1\ ,
\nonumber\\
|\Phi_{\rm boson}^{III}\rangle  &=& (a_{45}^\dagger)^{n_1} 
(a_{67}^\dagger)^{n_2} b_{-1/2}^{45} b_{-1/2}^{67}
b_{-3/2 }^{a}
\ |0\rangle _{\rm NS}\ ,\ \ \ 
a=45, \ 67\ ,\ \  n_1+n_2=\hat N-2\ .
\ee
These are $2(\hat N+1)+4\hat N+2(\hat N-1)=8\hat N$ bosonic states
for $\hat N=1,2,...$, and $2 $ bosonic states for $\hat N=0$.

\smallskip

\noindent R sector:
\be
|\Phi^I_{\rm fermion}\rangle &=& (a_{45}^\dagger)^{n_1} 
(a_{67}^\dagger)^{n_2}  |u\rangle _{\rm R}\ ,\qquad n_1+n_2=\hat N\ ,
\nonumber\\
|\Phi^{II}_{\rm fermion}\rangle &=& (a_{45}^\dagger)^{n_1} 
(a_{67}^\dagger)^{n_2} d_{-1}^a |u\rangle _{\rm R}\ ,
\qquad \ \  a=45, \ 67\ ,  \ \  n_1+n_2=\hat N-1\ ,
\nonumber\\
|\Phi^{III}_{\rm fermion}\rangle &=& (a_{45}^\dagger)^{n_1} 
(a_{67}^\dagger)^{n_2} d_{-1}^{45} d_{-1}^{67} |u\rangle _{\rm R}\  ,  \qquad  n_1+n_2=\hat N-2\ ,
\ee
where $|u\rangle _{\rm R}$ is the standard fermion vacuum of the
Ramond sector, with spin components aligned with the magnetic fields
$F_{45}$, $F_{67}$.
These are $2(\hat N+1)+4\hat N+2(\hat N-1)=8\hat N$ fermionic states,
with $\hat N=1,2,...$ (and 2 fermion states for   $\hat N=0$),
so in total there are $8\hat N + 8\hat N $ surviving states at each energy level
$\hat N\geq 1$.

\subsection{Non-supersymmetric string models}

One can break supersymmetry by  introducing  another magnetic field component
$F_{89}$.
The resulting spectrum is now
\beq
\a' M^ 2= \hat N-\vp (\hat J_1 + \hat J_2) - \vp_3 \hat J_3\ ,
\label{suno}
\eeq
\beq
\pi \varphi_3=\arctan 2\pi\a' e F_{89}+\arctan 2\pi\a' e' F_{89}\ ,
\label{stres}
\eeq
with $\hat J_3\equiv \hat J_{89}$. 
Now we scale the parameters as follows
\beq
\a' =\ep\ \a'_{\rm eff}\ ,\ \ \ \ \vp_3=\ep\ \bar\vp_3\ ,\ \ \ \ 
\vp=1-\ep\ ,
\eeq
and consider the limit $\ep\to 0$ with $\a'_{\rm eff}$ and $\bar\vp_3
$ 
fixed.
As a result, only states with maximum spin aligned with the magnetic
field remain of finite mass,
$ S_1+ S_2=\hat N+1$, or
$\hat J_1 + \hat J_2=\hat N$,
just as in the previous case. 
All other charged string states are decoupled.
The masses of the surviving states are now given by
\beq
\a'_{\rm eff} M^ 2= \hat N - \bar \vp_3 \hat J_3 \ .
\label{scua}
\eeq
There is a splitting between fermion and boson masses due to the
gyromagnetic coupling with the magnetic field $\bar \vp_3$.
In the NS sector, $\hat J_3=-l_3-\half \, ,\ l_3=0,1,2,...$, since the requirement $ S_1+
S_2=\hat N+1$ implies that $S_3=0$. This shows that this model has no tachyons.

\medskip

A different non-supersymmetric model can be obtained as follows.
One consider the model with $\vp_1 $ and $\vp_2$ and
write 
\beq
\a' =\ep\ \a'_{\rm eff}\ ,\ \ \ \ \ \vp_1=1-a_1 \ep\ ,
\ \ \ \ \ \vp_2=1-a_2 \ep\ .
\eeq
As $\ep\to 0$ with $ \a'_{\rm eff}$, $a_1$, $a_2$ fixed,
only the states with $\hat J_1 + \hat J_2=\hat N$ survive, just as before.
Their mass formula is given by 
\beq
\a'_{\rm eff} M^ 2= a_1  \hat J_1 +a_2  \hat J_2 \ , \ \ \ \ 
\hat  J_1 +  \hat J_2 =\hat N\ .
\label{scuad}
\eeq
{} For $a_1=a_2$ we recover the supersymmetric case.
The model is characterized by two independent parameters, say $\a'_{\rm eff}$
and $a_2$, since the other parameter, $a_1$, can be absorbed into a
redefinition of the $\a'_{\rm eff}$, $a_2$.

Unlike the previous non-supersymmetric model (\ref{scua}),
 this model has tachyons appearing for $\hat J_1=-\half $, $\hat
J_2=\hat N+\half $ or $\hat J_2=-\half $, $\hat
J_1=\hat N+\half $, with $\hat N<(a_1-a_2)/(2a_2)$ for $a_1>a_2$. 
They will be reflected as IR divergences in the
one-loop partition function.

\section{Partition function}
\setcounter{equation}{0}

As a further test of the model that results upon taking 
the limit of section 2, 
in this section we  compute the one-loop 
partition function.
This can be obtained by taking the limit of the partition function 
for open superstring theory in the presence of several magnetic fields
(at zero and finite temperature)  computed by Tseytlin in 
\cite{tseytlin}. 


\subsection{Non-supersymmetric models at zero temperature}

We first consider the string model with $\vp\equiv \vp_1=\vp_2$ and $\vp_3$.
In the present case, the partition function of \cite{tseytlin} takes the form
\beq
Z= c_1 \int_0^\infty {dt\over t^2}\ \prod_{I=1}^4  {f_I\over \theta_1(it\vp_I|it)}
\left[  \prod_{J=1}^4 \theta_2(it\vp_J|it) -\prod_{J=1}^4 \theta_3(it\vp_J|it)
+\prod_{J=1}^4 \theta_4(it\vp_J|it)\right]\ ,
\label{parti}
\eeq
$$
f\equiv f_1=f_2=2\pi\a'F_{45}(e +e')\ ,\ \ \ \
f_3=2\pi\a'F_{89}(e+e')\ ,\ \ \ \vp_4=0\ .
$$
Now we rescale parameters as follows:
\beq
\a'=\ep\, \a'_{\rm eff}\ ,\ \ \vp= 1-\ep\ ,\ \  \ ,\ \vp_3=\ep\bar\vp_3\ ,\ \ \ 
t={\t\over \ep} \ .
\eeq
Taking the limit $\ep\to 0$ with $\a'_{\rm eff},\ \bar\vp_3$ fixed,  
the partition function becomes
\beq
Z= \ep c_1f_0^2 \pi\bar\vp_3 \int_0^ \infty {d\t\over\t ^3}\,  \tanh({\pi \bar\vp_3 \t\over 2})\, {\big(
1-2 e^{-2\pi\t}\cosh(\pi\bar\vp_3 \t )+
e^{-4\pi\t}\big)\over (1-e^{-2\pi\t})^2}\ ,\ \ \ \ 
f_0={(e+e')^2\over \pi \, ee'}\ .
\label{gunt}
\eeq
For $\bar \vp_3=0$, the partition function vanishes identically, 
as expected, since the model becomes supersymmetric.
Note the overall $\ep $ factor.

The IR region is at $\t\to\infty $. The integral becomes $\int^\infty d\t/\t^3$
which is convergent. In the UV region, $\t\to 0$, and we find that
the integral diverges as  $\int_0\ d\t/\t^2$.

The result (\ref{gunt}) is very different  from what is obtained by a
direct $\a'\to 0$ limit of (\ref{parti}). In this limit
only the Yang-Mills supermultiplet survives, and --as observed in \cite{tseytlin}--
one finds a result which is proportional to the 
$d=10$ Yang-Mills one-loop effective action.
In the case of $SU(2)$ theory with a $U(1)$ background field 
$\hat F_{\mu\nu}=\sigma_3 F_{\mu\nu}$, with $F_{45}=f$, $F_{67}=f$ and
$F_{89}=f_3$, the one-loop effective action is \cite{chep,tse}
\beq
\Gamma_{\rm YM}=
\int_{\delta}^\infty {d\t\over\t^3} {f^2f_3\over \sinh^2(f\t)\sinh(f_3\t )}
\left(2\cosh(2f\t)+\cosh(2f_3\t)+1-4 \cosh^2(f\t)\cosh(f_3\t)\right)\ .
\eeq
 In the UV $\t\to 0$ region,  the integral behaves as  $\int_0\
 d\t/\t^2$,
just as (\ref{gunt}).

\medskip

Now consider the partition function of the non-supersymmetric
model with two magnetic fields and mass spectrum
given by (\ref{scuad}). We rescale parameters as follows
\beq
\a'=\ep\, \a'_{\rm eff}\ ,\ \ \vp_1= 1-a_1\ep\ ,\ \  \ ,\ \vp_2=1-a_2\ep\ ,\ \ \ 
t={\t\over \ep} \ .
\eeq
By taking the $\ep\to 0$ limit in (\ref{parti}), we obtain
\beq
Z= \half \ep c_1f_0^2  \int_0^ \infty 
{d\t\over\t ^4}\,  {\big( \cosh(a_1\pi \t)- \cosh(a_2\pi
  \t)\big)^2\over
\sinh(a_1\pi t)\sinh(a_2\pi t)}\ .
\eeq
In the UV $\t\to 0$ region,  the integral behaves as  $\int_0\
 d\t/\t^2$, as in the previous model.
In the IR $\t\to \infty $ region, the integral is divergent,
$\int^\infty \
 d\t/\t^4 e^{\pi (a_2-a_1) \t}$, for $a_2>a_1$. This is due to the
 presence of tachyons in this particular model as pointed out in section 2.

\subsection{Supersymmetric model at finite temperature}

Now we consider the model $\varphi\equiv \vp_1=\vp_2$ at finite
temperature 
$T=\beta^{-1}$.
The resulting model after taking the limit $\vp\to 1$
depends on three parameters $\a'_{\rm eff}$, $\beta$ and the string coupling $g_s$.

The string-theory partition function before the limit is given by \cite{tseytlin}
\beq
Z=a_1\beta \int_0^\infty {dt\over t^2}\, \theta_2\big(0|{i\beta^2\over 2\pi^2
\a' t}\big)\ \prod_{I=1}^4 f_I {\theta_2(it\hat
  \vp_I|it)\over\theta_1(it\vp_I|it)}\ .
\eeq
In the present case, $\vp_1=\vp_2\equiv \vp $, 
$\vp_3=\vp_4=0$, $\hat\vp_1=\hat\vp_2=0$,
$\hat\vp_3=\hat\vp_4=\vp $~.

Now rescale variables as follows
\beq
\a'=\ep\, \a'_{\rm eff}\ , \ \ \ \vp=1-\ep\ ,\ \ \ \ \ t={\t\over\ep
}\ .
\eeq
After taking the limit $\ep\to 0$ with fixed $\a'_{\rm eff},\ \beta $, 
we find the following partition function
\beq
Z=\ep  a_1f_0^2 \beta  \int_0^\infty {d\t\over\t^4}\, 
\theta_2\big(0|{i\beta^2\over 2\pi^2
\a'_{\rm eff} \t}\big)\ {(1+e^{-2\pi\t})^2\over (1-e^{-2\pi t})^2 }\ .
\label{ther}
\eeq
This is in agreement with the general expression for the free energy in $p+1$ dimensional field theory
with mass operator $\hat M=\hat M_B=\hat M_F$\ ,
\beq
Z=V_p \beta  \int_0^\infty {d\t\over\t^{p+3\over 2}}\ \theta_2\big(0|{i\beta^2\over 2\pi^2
\a'_{\rm eff} \t}\big)\ {\rm Tr}\ e^{-\t M^ 2}\ .
\eeq
To see this, we note that, in the present case, the boson and fermion
mass spectrum of section 2 is given by $\alpha'_{\rm eff}\hat M^2=\hat
N$, with $\hat J_1+\hat J_2=\hat N$, 
and $p=5$ is the number of
spatial dimensions with translational invariance (i.e. $x^\mu$ with
$\mu=1,2,3,8,9$).

Let us now examine the convergence properties of the thermal
partition function.
\smallskip

\noindent -- In the UV limit, $\t\to 0$ and $\theta_2\big(0|{i\beta^2\over
  2\pi^2\a'_{\rm eff} \t}\big)\to 2\exp\big[ - {\beta^2\over
  8\pi \a'_{\rm eff} \t}\big]$ and the integral is convergent
for any value of $\beta $. This shows that there is no Hagedorn
critical temperature. This is expected, since the Hagedorn temperature
is due to the rapid growth of the density of string states with the
  mass. In the present case, most of the string states have decoupled.
The growth of the number of surviving string states is only linear  with 
$\hat N $ and as a result no Hagedorn transition appears.

\smallskip

\noindent -- In the IR region,  $\t\to \infty $. Using the modular transformation
$\theta_2(0|{i\over \tau})=\sqrt{\tau} \theta_4(0|{i \tau})$, and
the fact that $ \theta_4(0|{i \tau})\to 1$ as $\tau\to\infty $, we
find that the integral (\ref{ther}) takes the form $\int^\infty
d\t/\t^{7/2} $,
which is convergent at infinity.

\section{Type II superstrings}
\setcounter{equation}{0}


A simple model of type II string theory in a magnetic background
is the one considered in \cite{RT1}, where the magnetic field is of Kaluza-Klein origin. 
This has been generalized to include several magnetic field components
in \cite{RT2} and in \cite{TU}.
We follow the notation of \cite{RT2}.
The bosonic part of the GS lagrangian is
\be
L_B &=&\p_+x_i\p_-x_i+ (\p_+ + ib_1 \p_+ y) z_1 (\p_- - ib_1 \p_- y) z_1^*
\non\\
&+& (\p_+ + ib_2 \p_+ y) z_2 (\p_- - ib_2 \p_- y) z_2^* + \p_+ y \p_-
y\ ,
\ee
$$
z_1=x_4+ix_5\ ,\ \ \ z_2=x_6+ix_7\ .
$$
Here $y$ is a compact coordinate with $y=y+2\pi R$.
This is an exact CFT. This is clear from the fact that the $\sigma
$-model metric 
describes Minkowski space with some identifications, so the Riemann tensor is identically zero.
The dimensional reduction in $y$ gives a magnetic background with curvature proportional
to the electromagnetic stress tensor. Thus taking a strong magnetic limit will imply strong 
curvatures for the dimensionally reduced metric.

We consider a  configuration with $b_1=b_2$, which preserves 1/2 of
the original 32 supersymmetries \cite{RT2,TU}.
The superstring spectrum is given by
\beq
\a'M^2=2\big(\hat N_R-\hat \g (\hat J_{1R} +\hat  J_{2R})\big)+
2\big(\hat N_L+ \hat \g (\hat J_{1L} +\hat  J_{2L})\big)+{\a'\over R^2}
\big( m-bR(\hat J_1+\hat J_2)\big)^2+{w^2R^2\over \a'}
\eeq
\beq
\hat N_R-\hat N_L=mw\ ,\ 
\label{match}
\eeq
\beq
\hat J_s=\hat J_{sL}+\hat J_{sR}\ ,\ \ \ \  
\hat J_{sL}=S_{sL} +l_{sL}+\half \ ,\ \ \ \ 
\hat J_{sR}= S_{sR} -l_{sR}-\half\ ,\ \ \ s=1,2\ ,
\eeq
$$
l_{sL},\ l_{sR}=0,1,2,...
$$
where $m$ and $w$ are, respectively, 
Kaluza-Klein momentum and winding numbers around the compact
direction  $y$,  $\hat\g\equiv \gamma- [\gamma ]$,
 $\g=wbR $, 
and $[\g ]$ is the integer part of $\gamma $.

Now we write
\beq
\a '=\ep\, \a'_{\rm eff}\ , \ \ \ \ bR=1-\ep\ ,\ \ \ \ R=\ep \bar R \ ,
\eeq
so that $\hat \g =1-w\ep $ for $w\neq 0$.
We find
\be
\a'_{\rm eff}M^2 &=& {2\over\ep}(\hat N_R-(1-w\ep )(\hat J_{1R} +\hat  J_{2R})\big)+
{2\over \ep}\big(\hat N_L+(1-w\ep) (\hat J_{1L} +\hat  J_{2L})\big)
\non\\
&+& {\a' _{\rm
    eff}\over
\ep^2\bar R^2}
\big( m-(1-\ep) (\hat J_1+\hat J_2)\big)^2+{w^2\bar R^2\over \a'_{\rm
    eff}}\ .
\ee
Therefore the only $w\neq 0$ states which have finite mass in the
limit $\ep\to 0$ with $\a'_{\rm eff},\ \bar R$ fixed are those with the
following
quantum numbers
$$
\hat J_{1R} +\hat  J_{2R}=\hat N_R\ ,\ \ \ \hat J_{1L} +\hat  J_{2L}=-\hat N_L\ ,
$$
\beq 
m=\hat J_1+\hat J_2=\hat N_R-\hat N_L\ ,\ \ \ \ w=1\ .
\label{qunu}
\eeq
They have finite masses given by 
\beq
\a'_{\rm eff}M^2=2  (\hat N_R+\hat N_L) +{\a' _{\rm
    eff}\over\bar R^2}
\big(  \hat N_R -\hat N_L  \big)^2+{\bar R^2\over \a'_{\rm eff}}\ .
\label{sugma}
\eeq

Now consider the neutral sector with $w=0$. In this case the spectrum
is
the same as the free superstring spectrum in flat Minkowski spacetime
\cite{RT1}.
Therefore
\beq
\a'_{\rm eff}M^2= {2\over\epsilon}( \hat N_R+\hat N_L )\ ,\ \ \ \ 
 \hat N_R=\hat N_L \ .
\eeq
In the limit $\ep\to 0$, all string excitations  decouple, leaving 
only the massless supergravity multiplet
$\hat N_R=\hat N_L=0$.

Thus the full theory after the limit is type II supergravity coupled to
the infinite number of states (\ref{qunu}) of the leading Regge trajectory
with masses given by  (\ref{sugma}).


\bigskip

To conclude, we have seen that there is a limit in superstring theory
in
which
all string excitations can be decoupled except for
certain string states lying on the leading Regge trajectory.
This implies a great simplification of the theory. 
An interesting problem is to understand if the resulting theory
can be described in terms of a (Lorentz invariant)
quantum field theory in 3+1 dimensions.
It would be interesting to see if it is possible to
implement this limit  at the level of
string  sigma model, and to see how to formulate
interactions
in the resulting model without having to go through a limiting
process of string-theory scattering amplitudes.
It would also be interesting to see if one can take 
analogous limits in Dp-branes.





\section{Acknowledgements}

I would like to thank A. Tseytlin for useful comments,
and M. Sheikh-Jabbari for a useful discussion.
This work is partially supported by  MCYT FPA
2004-04582-C02-01, the EC-RTN network
MRTN-CT-2004-005104 and by CIRIT GC 2001SGR-00065.

{}  

\end{document}